\documentclass{aastex631}

\usepackage{subfigure}
\usepackage{hyperref}
\usepackage{amsmath}
\usepackage{graphicx}
\usepackage{xcolor}
\usepackage{diagbox}
\usepackage{tabularx}
\usepackage{booktabs}

\newcommand{\minute}{\ensuremath{^\prime}}

\begin{document}

\title{SNR G54.1+0.3, a PeVatron candidate unveiled by LHAASO}

\correspondingauthor{Yudong Cui, Lili Yang}
\email{cuiyd@mail.sysu.edu.cn, yanglli5@mail.sysu.edu.cn}

\author{Yihan Shi}
\author{Yudong Cui}
\affiliation{School of Physics and Astronomy, Sun Yat-sen University, No. 2 Daxue Road, 519082, Zhuhai China}

\author{Lili Yang}
\affiliation{School of Physics and Astronomy, Sun Yat-sen University, No. 2 Daxue Road, 519082, Zhuhai China}
\affiliation{Centre for Astro-Particle Physics, University of Johannesburg, P.O. Box 524, Auckland Park 2006, South Africa}
\begin{abstract}

Recently, the LHAASO Collaboration reported the first very-high-energy gamma-ray catalog, containing 90 TeV sources. Among these sources, 1LHAASO J1929+1846u is located 0.3$^\circ$ west of SNR G54.1+0.3 and also lies within a $+53 \, \text{km s}^{-1}$ cloud (the Western Cloud).
Moreover, one of the IceCube track-type high-energy starting events is found around 1.3$^\circ$ north of 1LHAASO J1929+1846u, which may serve as strong evidence for the hadronic origin of this TeV source.
SNR G54.1+0.3 is a young supernova remnant (SNR), with a powerful pulsar wind nebula (PWN) inside. Its X-ray radiation from the PWN and the SNR Shell can be clearly identified. The radio emission from the PWN region is also given. However, given the angular resolution of gamma-ray experiments, the entire SNR region is viewed as a point source by Fermi-LAT, H.E.S.S. and VERITAS. In this work, we explore a hybrid scenario where SNR G54.1+0.3 is indeed associated with the Western Cloud, and we derive the multi-wavelength emissions from the PWN, the SNR Shell, and the Western Cloud, separately. Our model can explain the observations well, indicating that SNR G54.1+0.3 might be an excellent candidate of Galactic PeVatron and neutrino source.
\end{abstract}


\vspace{2\baselineskip} 
\section{Introduction} \label{sec:intro}

The origin of cosmic rays (CRs) has been a longstanding question in the field. CRs with energies above $10^{18} eV$ are commonly believed to originate from extragalactic sources, such as Active Galactic Nuclei (AGN) and Gamma-Ray Bursts (GRBs). CRs with lower energies may come from Galactic sources, such as pulsars and Supernova Remnants (SNRs).
PeVatrons, as sources where cosmic rays are accelerated to petaelectronvolt (PeV) energies in the Milky Way, are believed to be the reason for the knee feature of the cosmic ray spectrum. 
CRs primarily gain energies via diffusive shock acceleration at shock waves.
Energetic CRs will collide with ambient matter and radiation through $pp$ and $p\gamma$ interactions within their source regions. Consequently, high-energy neutral gamma rays and neutrinos are generated \citep{Stecker:1978ah, Kelner:2008ke}. To identify the origin of Galactic CRs, multi-wavelength and multi-messenger observations and studies would be the best method and have been extensively performed in recent years \citep{Gonzalez-Garcia:2009bev,icecube2018multimessenger, Sarmah:2023pld, Sudoh:2022sdk}. 

Currently, the most studied PeVatron candidates include pulsar wind nebulae (PWNe) \citep{burgess2022eel}, supernova remnants (SNRs) \citep{enomoto2002acceleration, zhang2019supernova, tibet2021potential} and Galactic center regions \citep{Albert:2024aaa, Scherer:2022gft}. However, none of them has been definitively confirmed, because their gamma-ray emission can be generated both through the leptonic process of the inverse Compton (IC) scattering and through the hadronic process of proton interactions. During hadronic processes, neutrinos as counterparts can be generated, which can serve as a smoking gun for proton acceleration. As the IceCube neutrino observatory has been running for more than a decade, a large amount of data has been accumulated. The joint analysis and correlation search has been carried out\citep{ANTARES:2015moa,IceCube:2015afa,HAWC:2024kkc}.  

Special thanks to the successful operation of the very-high-energy (VHE) ground-based gamma-ray experiment , LHAASO, about 90 Galactic sources with emissions above TeV energies have been observed \citep{cao2024first}. These groundbreaking discoveries have opened a new window to gamma-ray astronomy and present a step forward in the search for the PeVatrons. Within the first LHAASO catalog, we found one high-energy starting event (HESE) neutrino observed by the IceCube experiment associated with 1LHAASO J1929+1846u. Moreover, this LHAASO source is located 0.3$^\circ$ west of SNR G54.1+0.3, and lies within a $+53 \, \text{km s}^{-1}$ cloud (the Western Cloud). SNR G54.1+0.3 was first discovered at a frequency of 4.75 GHz, and this radio source was suggested as a Crab-like SNR for its flat spectral index and filled-center morphology \citep{reich1985evidence, velusamy1988g54}. It is relatively small, with a size of 1.2 \minute and contains a pulsar J1930+1852 with a period of 136 ms and an age of about 2900 years \citep{camilo2002discovery, chevalier2005young,bocchino2010xmm,torres2014time,gelfand2015properties,leahy2018evolutionary}. 
Radio, infrared (IR) and X-ray emissions of PWN were detected by Very Large Array (VLA) \citep{lang2010radio}, AKARI IRC, SOFIA FORCAST, Spitzer MIPS, Hersche PACS, Hersche SPIRE \citep{temim2017massive}, XMM-Newton and SUZAKU \citep{bocchino2010xmm}, respectively. 
Its X-ray shell has been detected around the pulsar with a radius r of $ 160''-400''$ \citep{bocchino2010xmm}. This is a typical composite system where the PWN is surrounded by a shell-like SNR \citep{gaensler2006evolution}. 
Gamma-ray experiments, like the High Energy Stereoscopic System (H.E.S.S.) \citep{collaboration2018hess}, VERITAS \citep{abeysekara2018veritas} and Fermi-LAT also detect the signal excess from PWN/SNR, but are not bale to resolve the shell structure from the PWN clearly. The LHAASO observation around this SNR covers a much larger area, hence its flux is much higher than that of VERITAS, and its center location lies close to the Western Cloud rather than the SNR. 

The previous study \citep{li2010lepto} has investigated the origin of gamma-ray emissions from G54.1+0.3. They argued that a pure leptonic model fails to simultaneously explain the observations in both X-ray and gamma-ray bands, whereas a hybrid model incorporating both leptonic and hadronic scenarios would resolve the issue effectively. 
This type of model is not uncommon, such as another famous likely hybrid model source G106.3+2.7 \citep{fujita2021x}.  
G54.1+0.3 is young and its current shell is still very powerful, which may be capable of accelerating particles to PeV energies. 
Therefore in this work we use the leptonic emission from the PWN and the SNR Shell to explain the radio, X-ray, and gamma-ray observations, and by adding the hadronic interaction of the escaped CRs from the SNR, our hybrid model can explain the LHAASO observations at/around the Western Cloud.

We organize the manuscript as follows. In Section \ref{sec:data}, we present the multi-messenger observations, our Fermi-LAT data analysis, and the derived physical parameters of the SNR, PWN and clouds accordingly. In Section \ref{sec:model}, we give our hybrid models which explain the multi-messenger observation for these three parts respectively. In Section \ref{sec:results}, discussion and conclusion are given.

\vspace{2\baselineskip} 
\section{Multi-messenger observations} \label{sec:data}

\subsection{Radio, Infrared, and X-Ray Observations}\label{2.1}

In 2010, \cite{bocchino2010xmm} presented the XMM-Newton and SUZAKU's X-ray observations of the PWN G54.1+0.3. They showed the PWN emission $r\lesssim 160''$ and the diffuse shell emission $r\sim 160''-400''$ . 
The SNR Shell region can be modeled by a power law function with a photon index of $\gamma$= 2.9, while $\gamma$= 1.82 for the PWN region \cite{bocchino2010xmm}.
Infrared detection mostly concentrated inside r$<4'$ was given by multiple detectors, such as AKARI IRC, SOFIA FORCAST, Spitzer MIPS, and Herschel Space Observatory \citep{temim2017massive}. The IR emission is mostly thermal radiation, but it provides an upper limit for the non-thermal radiation in our work.
The VLA has also conducted multi-frequency radio studies of the PWN G54.1+0.3 \citep{lang2010radio}. Its high-resolution observations reveal that G54.1+0.3 has a complex structure, including magnetized filaments and rings. At 1.4 GHz, a diameter of $\sim8'$  shell can be seen, however, due to the strong background, the emission from the Shell region is not given \citep{lang2010radio}.  
For the entire region ($\sim2'$  $\times1.'5$ ), the G54.1+0.3 spectrum obtained from the integrated magnetic flux density, has the best fit spectral index of $\alpha$ = $-0.28$.  The authors of \cite{lang2010radio} notice that the radio profile of the PWN is different from its X-ray profile. That may indicate that the radio and X-ray emission come from two groups of electrons. Therefore, in our work a broken power law spectrum is introduced to represent the two groups of electrons.

\subsection{Fermi data analysis}\label{2.2}
\begin{figure}[htbp]
    \centering
    \includegraphics[width=0.49\textwidth]{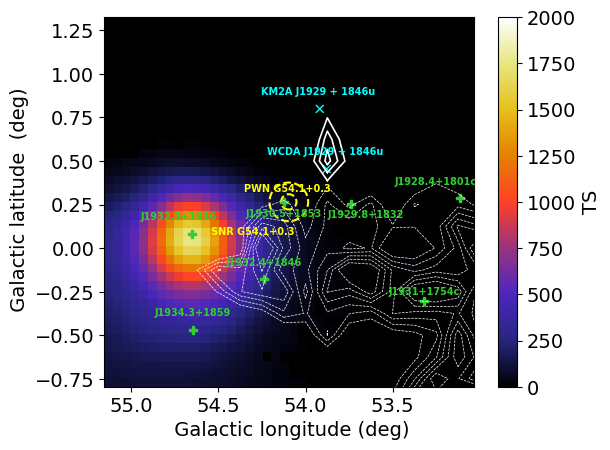}
    \hfill
    \includegraphics[width=0.47\textwidth]{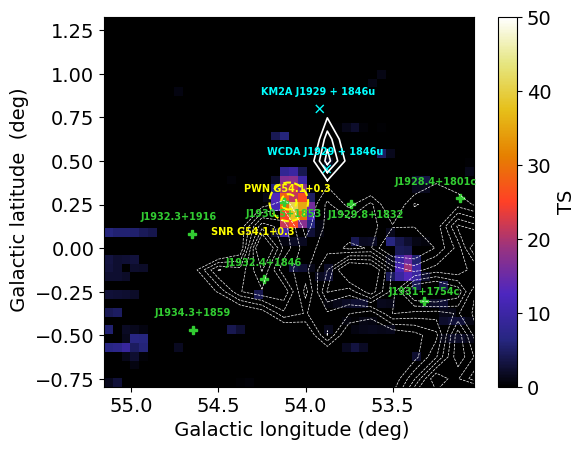}
    \caption{Fermi gamma-ray analysis in the ROI. In the left panel, we show the point sources (TS$>$10) in this ROI including J1932.3+1916, J1934.3+1859, J1932.4+1846, J1930.5+1853, J1929.8+1832, J1928.4+1801c, J1931+1754c, these 4FGL sources are represented as green plus symbols. In the right panel, we show the TS map after deducting all the point sources except J1930.5+1853. In both panels, the cyan cross represents the LHAASO source J1929+1846u, the yellow dashed circle represents the SNR G54.1+0.3, and the smaller dashed circle represents the PWN G54.1+0.3, the white thin lines are significance contours of the molecular clouds from 3.5 K to 12.5 K, in steps of 1.5 K, with integration speed ranging from $+50.7 \, \text{km s}^{-1}$ to $+59.8 \, \text{km s}^{-1}$, here we use thicker lines to represent the Western Cloud. }
    \label{fig1}
\end{figure}

Moreover, gamma rays were detected by Fermi-LAT, VERITAS \citep{abeysekara2018veritas} and H.E.S.S. \citep{collaboration2018hess} in the SNR region, but they were unable to distinguish the Shell from the PWN, due to the low angular resolution.
Here we download the Fermi data from the website\footnote{\url{https://fermi.gsfc.nasa.gov/ssc/data/access}}.
The time window is from October 1st 2010 to October 1st 2023 and the searching radius is about 1.5 degrees around the PWN.
We find some sources within the region of interest (ROI), which are J1932.3+1916, J1934.3+1859, J1932.4+1846, J1930.5+1853, J1829.8+1832, J1928.4+1801c, J1931+1754c as labeled in Figure \ref{fig1}. Among them, J1930.5+1853 is closest to SNR G54.1+0.3. We analyze the data with the application of Fermitools, a dedicated toolkit provided by the Fermi Collaboration. We first perform a binned likelihood analysis. With the commands gtselect, gtmktime to select and filter the data, we subtract the diffuse backgrounds and point source contribution with the galactic, extragalactic diffuse emission model (gll\_iem\_v07, iso\_P8R3\_SOURCE\_V3\_v1) and the 4FGL catalog.

The Binned Analysis library and pyLikelihood module are imported to run the likelihood analysis through the raw model file created by the LATSourceModel package. We get the model file that contains the fitted model for the whole ROI.  
By putting the fitted models of all the other point sources into the background model file, we calculate the significance of J1930.5+1853 by likelihood analysis. Finally, we obtain the test statistic (TS) maps around the LHAASO source J1929+1846u, by using the TSMap function in the gt\_apps package. The morphology and significance of the sources are shown in Figure \ref{fig1}.
By binned likelihood analysis, we obtained the spectrum energy distribution (SED) of J1930.5+1853. 

\vspace{1\baselineskip} 
\subsection{LHAASO observation}\label{2.3}
The Large High Altitude Air Shower Observatory (LHAASO) is a gamma-ray and cosmic-ray observatory located in Daocheng, Sichuan Province, China \citep{LHAASO:2019qwt, LHAASO:2021gok}. It is a multi-purpose integrated extensive air show (EAS) array consisting of three components, which are Square Kilometer Array (KM2A), Water Cherenkov Detector Array (WCDA) and Wide-Field Air Cherenkov Telescope Array (WFCTA), designed to study CR and gamma rays in a wide energy range from sub-TeV to more than 1 PeV. WCDA mainly detects 1 to $\sim$ 10 TeV, while KM2A is on the order of 100 TeV.

\begin{figure}[htbp]
    \centering
    \includegraphics[width=0.51\textwidth]{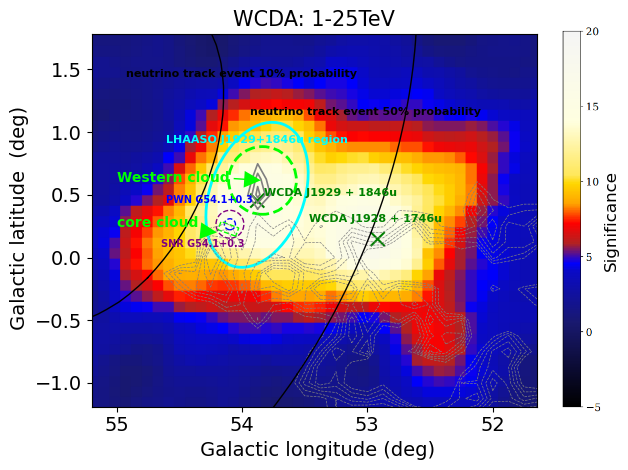}
    \hfill 
    \includegraphics[width=0.49\textwidth]{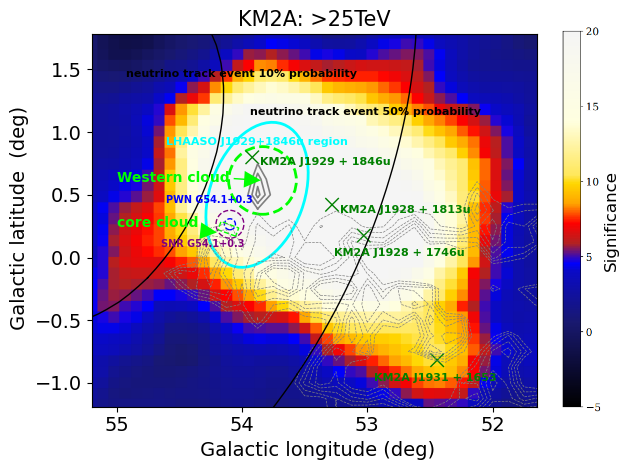}
    \hfill
    \includegraphics[width=0.49\textwidth]{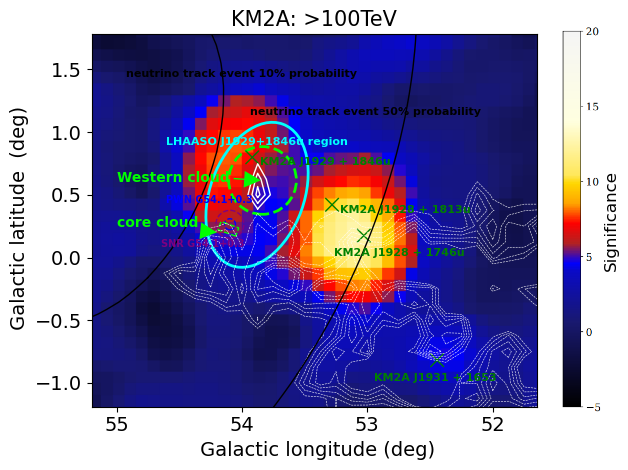}
    \caption{The three significance maps are get from Figure \textcolor{blue}{8} of \cite{cao2024first}, maps for energies lower than 100 TeV are saturated in/around 1LHAASO J1929+1846u. The green cross symbol represents the LHAASO source, and the gray dashed line shows the molecular clouds profile. The cyan ellipse roughly represents the WCDA J1929+1846u region, which covers both SNR G54.1+0.3 and the Western Cloud. The light green dotted circle represents the Western Cloud region, and the light green dotted rectangle represents the Core Cloud region. The blue dotted circle indicates the PWN region, and the purple dotted circle indicates the Shell region \citep{bocchino2010xmm}. Two black solid lines represent the 10\% and 50\% probability contours of the neutrino event, respectively.}
    \label{fig2}
\end{figure}

Recently, LHAASO announced its first catalog, containing about 90 VHE gamma-ray sources \citep{cao2024first}. One of the sources, named 1LHAASO J1929+1846u, was detected by both WCDA and KM2A. To distinguish between them, we refer to them as WCDA J1929+1846u and KM2A J1929+1846u, respectively. WCDA J1929+1846u is centered at $\alpha_{2000} = 292.34$, $\delta_{2000} = 18.77 $, and KM2A J1929 + 1846u is centered at $\alpha_{2000} = 292.04$ , $\delta_{2000} = 18.97 $.  Of the two, WCDA J1929+1846u is particularly close to the SNR G54.1+0.3. As seen in Figure \ref{fig2}, 1LHAASO J1929+1846u covers a wide area containing no less than the SNR region and the Western Cloud.  Meanwhile, two nearby sources, 1LHAASO J1928+1746u and J1928+1813u, are located much further from the SNR, hence are not discussed in this work.

The image shown in Figure \ref{fig2} is obtained from Figure 8 of \cite{cao2024first}, which is clearly saturated in/around the SNR region, no detailed structure is revealed.
Via private communication with the LHAASO team, we notice that both the WCDA and KM2A images tend to have a clear two-blob structure, with the northern blob centered at the Western Cloud and the southern blob centered at J1928+1746u. Most interestingly, the emission of northern blob in KM2A is mostly concentrated inside the Western Cloud region, i.e., the green dashed circle, meanwhile, the emission of the northern blob in WCDA extend more towards East and covers a region of both the SNR and the Western Cloud.  

This discovery seems to favor the scenario that the KM2A J1929+1846u is due to CRs escaped from the SNR illuminating the Western Cloud, and the WCDA J1929+1846u is caused by both the hadronic emission of the Western Cloud and the leptonic/hybrid emission of the SNR region.  

\vspace{2\baselineskip} 
\subsection{IceCube data }\label{2.4}

IceCube neutrino experiment has accumulated data for a decade, and a few high-energy neutrino candidates have emerged in the neutrino sky, such as NGC 1068 and TXS 0506+056 \citep{IceCube:2022der}. Although the galactic neutrino would contribute approximately 6-13 \% to the entire neutrino sky \citep{IceCube:2023ame}, no individual source has been identified. The simultaneous detection of neutrinos and gamma rays could confirm the presence of a galactic PeVatron. With the IceCube High Energy Starting Event (HESE) 12-year data sample, a selection of the highest energy neutrino whose interactions occurred within the detector fiducial volume \citep{IceCube:2020wum, IceCube:2023sov}, we perform a positional correlation search with LHAASO sources. There is a track event with energy of 67.6 TeV at RA=292.652, Dec=20.0527 observed on 21 February 2022. Its central position is 1.23 degrees from 1LHAASO J1929+1846u detected by KM2A as seen in Figure \ref{fig2}. It provides strong evidence for the hadronic origin of these VHE gamma-ray emission.

\vspace{2\baselineskip} 
\subsection{The distance of SNR G54.1+0.3 and its nearby clouds}\label{2.5}
The distance of SNR G54.1+0.3 and its association with nearby clouds are the key points in this work.
The work of \cite{leahy2008distance} applied the 21 cm absorption spectral line of HI to determine the lower limit of the distance of 4.5 kpc and the upper limit of 9 kpc. By detecting the emission spectrum of $^{13}$CO, \citep{leahy2008distance} found that PWN G54.1+0.3 appeared to be embedded in a molecular cloud at $+53 \, \text{km s}^{-1}$ (the Core Cloud), and this PWN is likely associated with this Core Cloud. We adapt the most updated rotation curve \citep{reid2019trigonometric}, and derive this PWN-cloud system at a distance of $\sim4.9$ kpc, which is consistent with the results of \citep{ranasinghe2018revised}. 
With the spectrum from ROSAT PSPC and ASCA, the authors of \cite{lu2001chandra} determined the distance of SNR G54.1+0.3 to be 5 kpc by measuring the X-ray absorption column density.
Here we adopt $4.9$ kpc as the distance of this SNR-cloud system.


In Figure \ref{fig2}, we show the outline of molecular clouds \citep{dame2001milky}. The Core Cloud is marked by a rectangle shape and the Western Cloud is highlighted by thick gray lines with a diamond shape. To calculate their mass, we get the the column density, N($H_2$), in unit of $cm^{-2}$ based on its relation with the integrated line intensity W(CO) and conversion factor $X_{\text{CO}}$ as in \citep{bolatto2013co}:
\begin{equation}
 N(\text{H}_2) = X_{\text{CO}} W(^{12}\text{C}^{16}\text{O}_{J=1 \rightarrow 0})
\label{eq:CO}
\end{equation}
The conversion factor is given by \( X_{\text{CO}} = 2 \times 10^{20} \, \text{cm}^{-2} (\text{K km s}^{-1})^{-1} \).
The mass of the Western Cloud calculated by integrating over the emission region and the velocity range from $+50.7 \, \text{km s}^{-1}$ to $+59.8 \, \text{km s}^{-1}$, is around \(1.9 \times 10^{4}\) $M_{\odot}$. The mass of the Core Cloud is about \(1.15 \times 10^{3}\) $M_{\odot}$.
As seen in Figure \ref{fig1}, the velocity has a large integral range, because the Western Cloud extends from $+50.7 \, \text{km s}^{-1}$ to $+59.8 \, \text{km s}^{-1}$. This is most likely due to strong turbulence. 
If we follow the velocity extension of the Core Cloud, which has an integral range from $+50.7 \, \text{km s}^{-1}$ to $+55.9 \, \text{km s}^{-1}$, the mass of the Western Cloud is about \(9.8 \times 10^{3}\) $M_{\odot}$. Here, we take the larger mass of \(1.9 \times 10^{4}\) $M_{\odot}$ for our study. With the mass of clouds, we can infer the approximate number of cold protons, which will be used to calculate the production of $pp$ interactions.

\vspace{2\baselineskip} 

\section{Models} \label{sec:model}
Accelerated charged particles emit electromagnetic radiation through a variety of radiative processes: synchrotron radiation, bremsstrahlung, and IC scattering for electrons, and for protons, decay into gamma rays or neutrinos of $\pi$ mesons produced in the interactions with the target proton or photons in the environment or the acceleration site.

In this section, we perform the modeling to the emission from the PWN, the Shell and the clouds to explain the multi-band observations. For the PWN region, the accelerated electrons can explain the radio to X-ray emission. However, they fail to explain the TeV to PeV gamma rays. For the SNR Shell, we use leptonic processes to explain the X-ray shell and GeV to Tev emission from the SNR G54.1+0.3. For the entire SNR region, we introduce the escaped CRs from SNR. Our model explains the LHAASO observations and gives the neutrino flux accordingly. 

\vspace{1\baselineskip} 
\subsection{PWN contribution}\label{3.1}
PWN G54.1+0.3 contains a rapidly rotating pulsar J1930+1852 \citep{camilo2002discovery}, its spin-down luminosity \( L = 1.18 \times 10^{37} \, \text{erg s}^{-1}\). 
Since the X-ray emission from the PWN can be separated from that of the Shell \citep{lu2001observations}, here we only consider the PWN contribution.  
The radio emission from the PWN region is given by \cite{lang2010radio}
GeV data from Fermi and TeV data from VERITAS represent the emission from the entire SNR region. 
LHAASO observation covering a much larger area including both the SNR and the Western Cloud is also shown in Figure \ref{fig2}.

We use GAMERA \citep{hahn2015gamera} to calculate the leptonic spectrum and assume that the injected electrons of PWN satisfy a broken power law distribution, which can explain most PWNe systems \citep{bucciantini2011modelling}, in terms of 
\begin{equation}
Q_{\mathrm{inj}}(\gamma, t) = 
\begin{cases} 
Q_0(t) \left( \frac{\gamma}{\gamma_b} \right)^{-\alpha_1} & \text{for } \gamma_{min} \leq \gamma \leq \gamma_b, \\
Q_0(t) \left( \frac{\gamma}{\gamma_b} \right)^{-\alpha_2} & \text{for } \gamma_b \leq \gamma \leq \gamma_{max},
\end{cases}
\label{eq:Q_inj}
\end{equation}

Where \(Q_0\) is the normalization factor depending on the total energy \(W_e\) of the PWN electrons, \(\alpha_1\), \(\alpha_2\) are the spectral indices, and \(\gamma\) is the Lorentz factor for relativistic electrons. 
The minimum Lorentz factor \(\gamma_{min} \sim 100\) (corresponding to electrons with energy \(\sim 50 MeV\)) to reproduce the radio emission, and maximum value \(\gamma_{max}\sim 2 \times 10^{8}\) (corresponding to electrons with energy \(\sim 0.1 PeV\)) .
The spectral indices are taken as $\alpha_1 = 1.7 $,  $\alpha_2 = 2.7$ and the cutoff energy is set to be $\gamma_b \cong 1 \times 10^{5}$ (\(\sim 50 GeV\)).

\begin{figure}[h]
    \centering
    \includegraphics[width=0.57\linewidth]{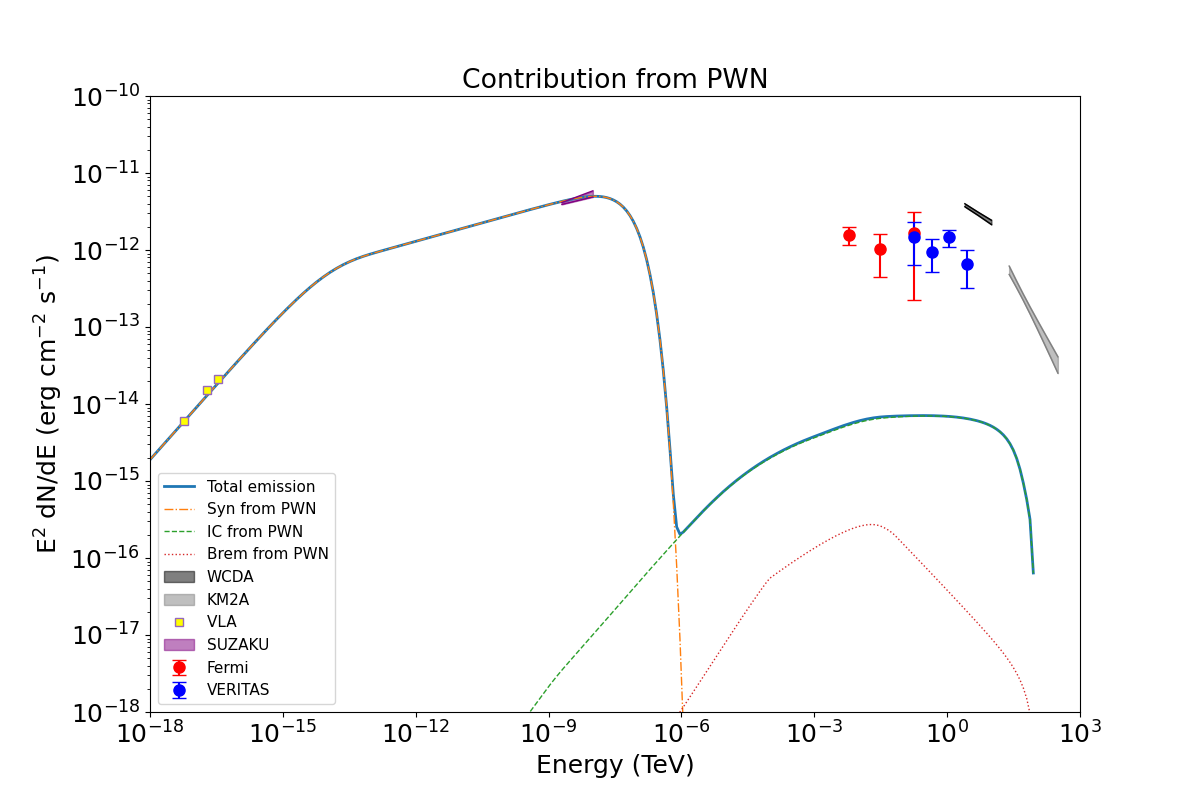}
    \caption{Modeling the leptonic emission of PWN G54.1+0.3. The X-ray and radio data  \citep{lang2010radio, bocchino2010xmm} are from the PWN region only.  Fermi (this work, red points) and VERITAS \citep{abeysekara2018veritas} data (blue points) are from the entire SNR region. WCDA and KM2A (gray shaded) emission cover a much larger region \citep{cao2024first} including both the SNR region and the Western Cloud. We use IC, synchrotron and bremsstrahlung in our model fitting.}
    \label{fig3}
\end{figure}

The leptonic emission includes three electron cooling mechanisms, which are synchrotron radiation, bremsstrahlung and IC scattering.
In the gamma-ray band, IC scattering makes the most prominent contribution.  
We set CMB temperature as 2.73 K, CMB energy density as 0.25 eV\,\text{cm}$^{-3}$, IR temperature 32 K, and IR energy density as 0.55 eV\,\text{cm}$^{-3}$.
The magnetic field strength is assumed to be 100 $\mu$G, which is also within the range of the work \citep{lang2010radio} 80 to 200 $\mu$G.
The calculated results are shown in Figure \ref{fig3}.

The fitting of our model shows good consistency with the radio and X-ray data, but in the high-energy band, the data clearly cannot be fitted. This finding agrees with the conclusions of the work by \cite{li2010lepto}, where pure leptonic model cannot explain the observations.  We will explore the contribution from the SNR Shell to solve this problem.


\vspace{2\baselineskip} 

\subsection{SNR shell contribution} \label{3.2}
In addition to the leptonic contribution from PWN, here we discuss the leptonic contribution from the SNR Shell. The high-energy observations shown here are the same as those in Section \ref{3.1}, but in the radio band, observations from the SNR Shell are not available due to strong background. For the X-ray band, we took the SNR Shell data from \citep{lu2001observations}.
In addition, we add the observations in the IR band \citep{temim2017massive}. Although this is the thermal radiation from nearby gas, it provides an upper limit of non-thermal radiation in our model.
We assume that the electrons follow a simple power-law distribution, with an exponential cut off to describe their maximum energy more precisely, in terms of 
\begin{equation}
 \frac{dN}{dE} = N_e \left( \frac{E}{1 \, \text{TeV}} \right)^{-\alpha_e}\exp\left(-\frac{E}{E_{cut}}\right)
\label{eq:dn/de}
\end{equation}

\begin{figure}[h]
    \centering
    \includegraphics[width=0.57\linewidth]{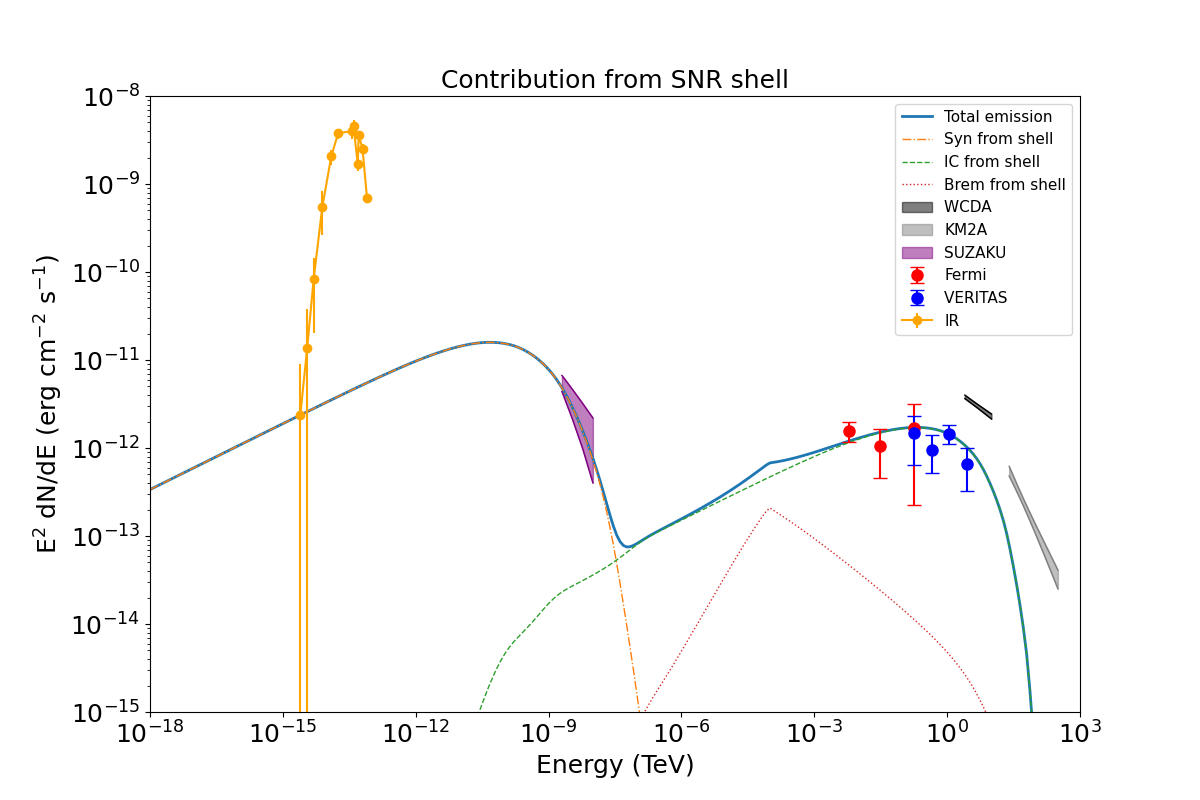}
    \caption{Modeling the leptonic emission of SNR G54.1+0.3. The observed data in X-ray \citep{bocchino2010xmm} (purple shaded) come from SNR Shell, IR data \citep{temim2017massive} (orange points) and $\gamma$-rays of Fermi (this work, red points), and VERITAS \citep{abeysekara2018veritas} (blue points) come from entire SNR region. WCDA and KM2A (gray shaded) emission cover a much larger region, including both the SNR region and the Western Cloud. We use IC, synchrotron and bremsstrahlung in our model fitting. }
    \label{fig4}
\end{figure}

Here \(\alpha_e\) and \(E_{cut}\) are the spectral index and cutoff energy. \(N_e\) is the normalization factor, depending on the total energy \(T_e\) . The relationship between \(N_e\) and \(T_e\) is
\begin{equation}
T_e = N_e \int_{E_{\text{min}}}^{E_{\text{max}}} E \left( \frac{E}{1 \, \text{TeV}} \right)^{-\alpha_e} \exp \left( -\frac{E}{E_{cut}} \right) \, dE.
\label{eq:Te}
\end{equation}
The best-fit parameters are $\alpha = 2.5 $,  \(T_e = 10^{49.4}\)erg and $E_{cut} \cong 19$ TeV. The magnetic field strength is assumed to be 15 $\mu$G. We have also used GAMERA \citep{hahn2015gamera} to calculate the leptonic emission, with the same radiation background settings as in Section \ref{3.1}.
The results are shown in Figure \ref{fig4}. We find that under this model, except for the observations of LHAASO, the low-energy and high-energy observations are fit well with the constructed model. The hadronic model is introduced in Section \ref{3.3} to explain the final piece, i.e., LHAASO emission.

\subsection{Molecular clouds contribution} \label{3.3}
To explain the discrepancy between the observation and calculation, we introduce hadronic emission from the Core Cloud and the Western Cloud. As $pp$ interactions produce ultra-high-energy gamma rays, but normally require a dense cloud to provide enough target protons. As mentioned in Section \ref{2.2}, the Western Cloud overlaps with 1LHASSO J1929+1846u. We also note a slight position offset between the WCDA and KM2A observations, where WCDA J1929+1846u covers both the SNR and the Western Cloud regions, while the KM2A J1829+1846u is more concentrated inside the Western Cloud.

We assume that the escaped CR has a simple power-law spectrum of the form
\begin{equation}
 \frac{dN}{dE} = N_p \left( \frac{E}{1 \, \text{TeV}} \right)^{-\alpha_p}\exp\left(-\frac{E}{E_c}\right),
\label{eq:pp SED}
\end{equation}
Where Np is the normalization factor, depending on the total energy of the injected protons, \(\alpha_p\)  and \(E_c\) are the spectral index and cutoff energy. The total CR released energy is $E_{total,CR}$.

In a homogeneous diffusion environment, we use Green's function to calculate the propagation process of escaped CRs,
\begin{equation}
 G = G(E, r, \Delta t) = \frac{1}{8(\pi \Delta t D)^{3/2}} \exp\left[-\frac{r^2}{4\Delta t D}\right], 
\label{eq:G}
\end{equation}

where D is the diffusion coefficient, r is the distance from SNR center to the cloud, and \(\Delta t\) is the propagation time. We assume an average propagation time for all CRs. At \(\Delta t\) ago, most of those CRs were instantly released and started to diffuse. The diffusion coefficient can be written as
\begin{equation}
 D = k D_{10} \left( \frac{E}{10 \, \text{GeV}} \right)^{b}, 
\label{eq:D}
\end{equation}
where \( D_{10} = 1 \times 10^{28} \, \text{cm}^2/\text{s} \), k is a coefficient of $D_{10}$, For more detailed discussions, refer to see \cite{berezinskii1990astrophysics} and \cite{ptuskin2006cosmic}. In addition, the masses of the clouds are given in Section \ref{2.3}. 

\begin{figure}
    \centering
    \subfigure[]
    {\includegraphics[width=0.51\textwidth]{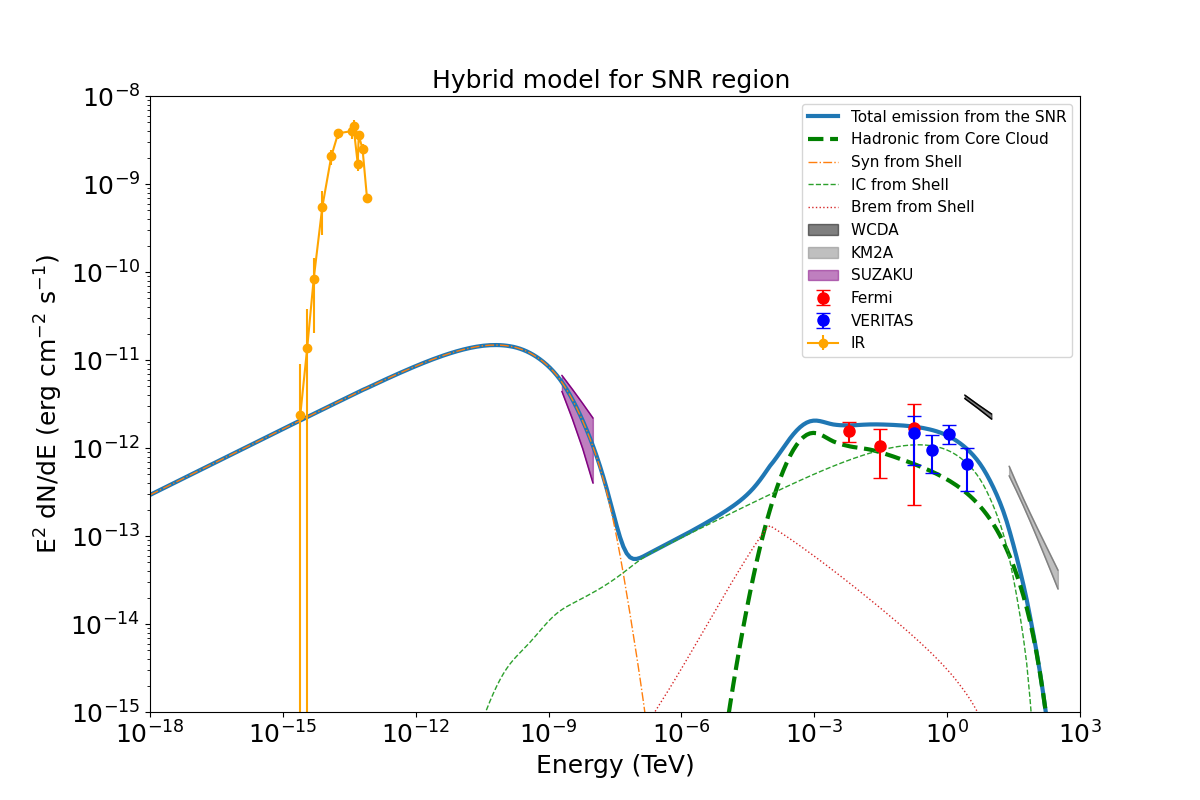}}
    \hfill
    \subfigure[]
    {\includegraphics[width=0.48\textwidth]{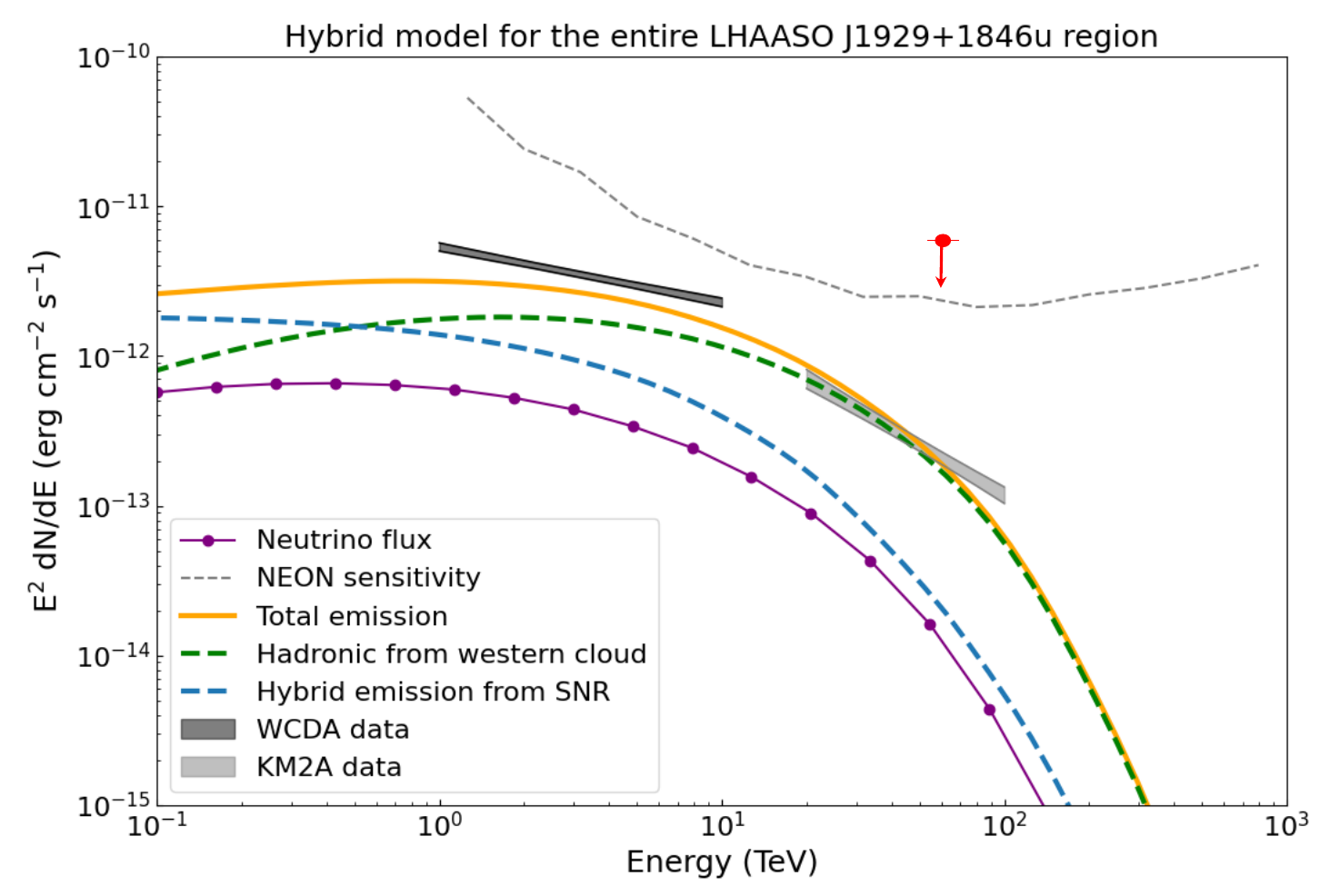}}
    \caption{Left panel: the hybrid emission model for the SNR region including the Shell and the Core Cloud. The leptonic process in the Shell region including synchrotron, bremsstrahlung and IC scattering are presented. The hadronic process in the Core cloud is shown with green dashed line. Right panel: the hybrid model for entire 1LHAASO J1929+1846u region. The emission from the SNR region is labeled as dashed blue line. Hadronic contribution from the Western Cloud is shown as green dashed line. The sum of these two, the total emission, is presented with orange solid line. The predicted neutrino flux and IceCube limit are shown as magenta line and red point.  The The observed data in X-rays \citep{bocchino2010xmm} come from SNR shell, infrared data \citep{temim2017massive} and $\gamma$-rays (Fermi, VERITAS \citep{abeysekara2018veritas})  come from entire SNR region. WCDA and KM2A emission cover a much larger region, including both the SNR region and the Western Cloud. The radiation processes include IC, synchrotron, bremsstrahlung and $pp$ interaction.}
    \label{fig5}
\end{figure}

There are four fitting parameters in our model, \(N_p\), \(\alpha_p\), \(E_c\) and \(k\). We fit both the KM2A SED and WCDA SED of J1929+1846u \citep{cao2024first} by using the MCMC method. 
The final fitting result can be seen in Figure \ref{fig5}. We obtained CR spectral parameters as follows: \(N_p = 8.35 \times 10^{48} erg^{-1}\), \(\alpha_p=1.65\) , \(E_c=154\, \text{TeV}\). The best-fit value of the diffusion coefficient is \(k = 0.405\) \(b = 0.38\), the propagation time is set to 1500 yr (where 1500 yr represents an average propagation value for a simplified model). To enhance the 1 - 25 TeV emission, the diffusion distances are set to be the projected distances, which are \(\sim 37\) pc for the Western Cloud, and \(\sim 0\) pc for the Core Cloud. The total energy of released CRs is also set relatively high as \(2 \times 10^{50} erg\).

The most difficult part in our model is to match the flux of WCDA J1929+1846u, hence, we adopt a relatively high total CR energy of \(2 \times 10^{50} \) erg, the shortest distance 37 pc between the Western Cloud and the SNR.  Additionally, the WCDA J1929+1846u flux seems too high to match the KM2A J1929+1846u flux. Two possibilities are as follows:
\begin{itemize}
    \item Firstly, the LHAASO data shown in the SED plot is only with statistic uncertainty, taking into account a systematic uncertainty of $8\%$, our model is consistent with the data.
    \item Secondly, due to the analysis method, the flux of WCDA J1929+1846u may be contaminated by WCDA J1928+1746u in the LHAASO data analysis. We expect WCDA background reduction to be improved.
\end{itemize}

\begin{figure}[htbp]
    \centering
    \subfigure[]
    {\includegraphics[width=0.48\textwidth]{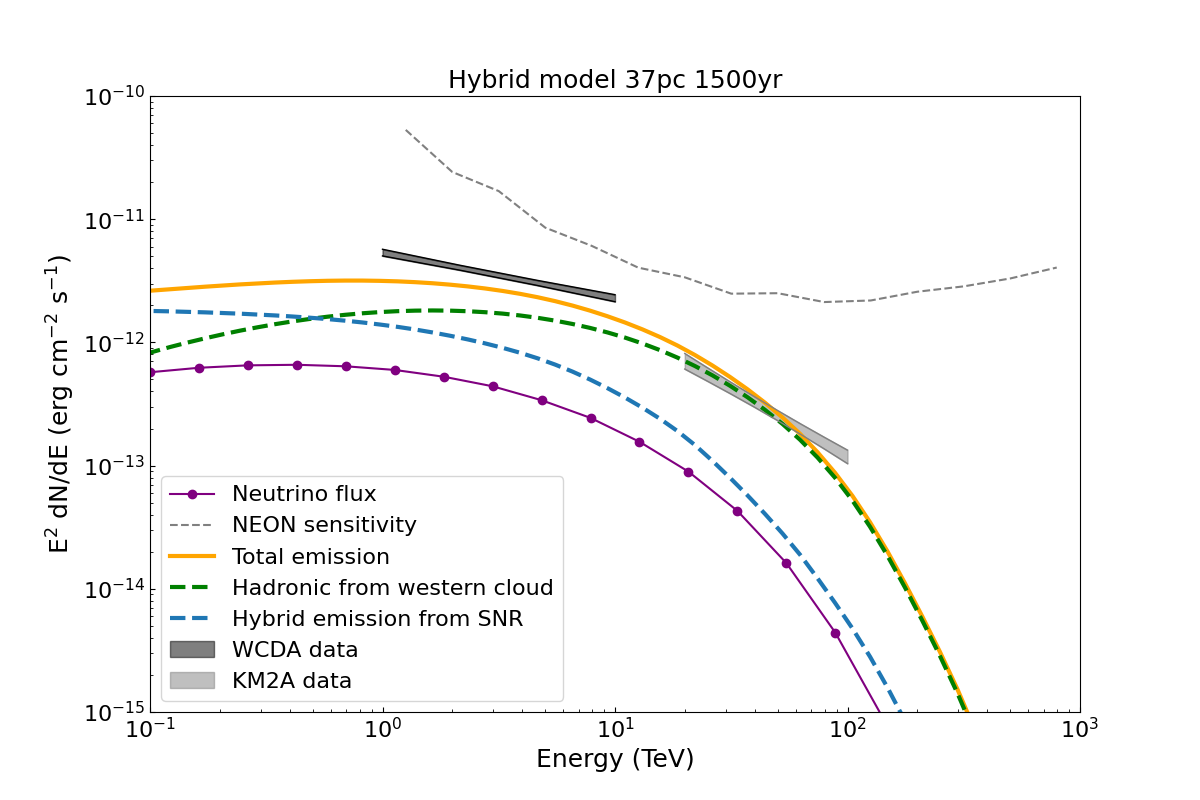}}
    \hfill
    \subfigure[]
    {\includegraphics[width=0.48\textwidth]{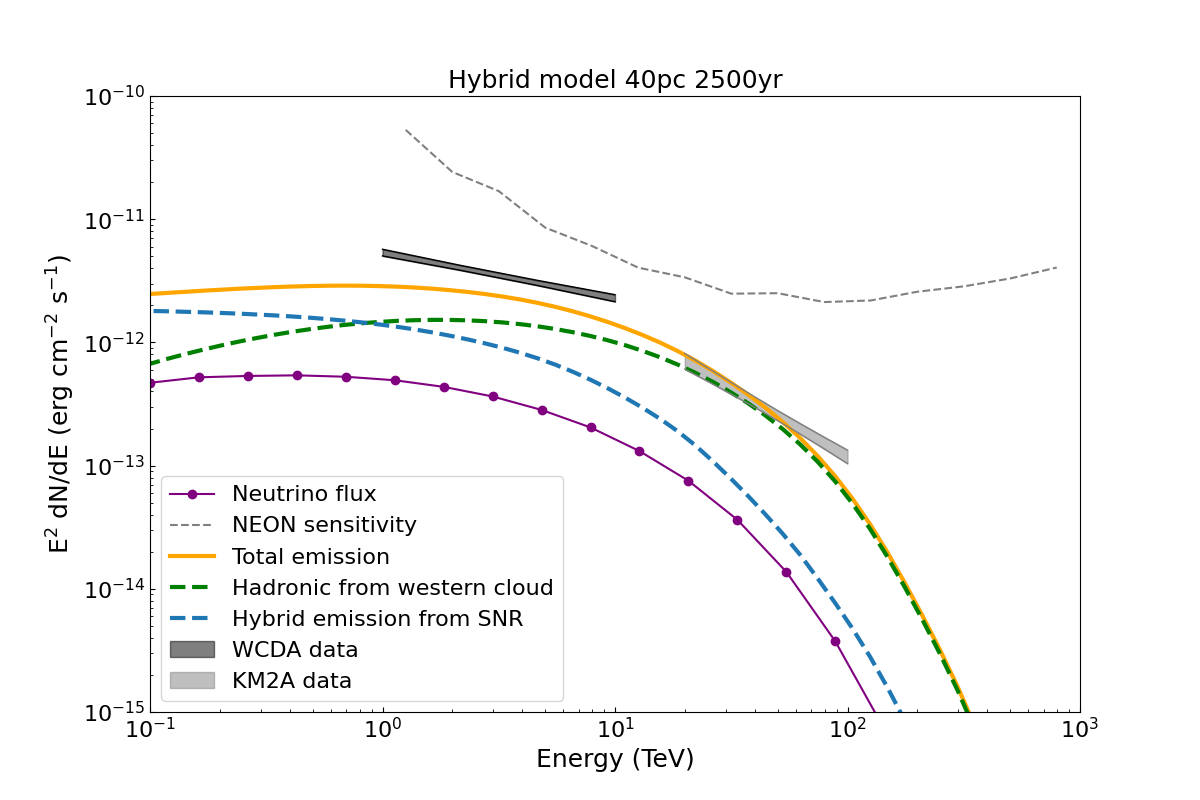}}
    \\
    \subfigure[]
    {\includegraphics[width=0.48\textwidth]{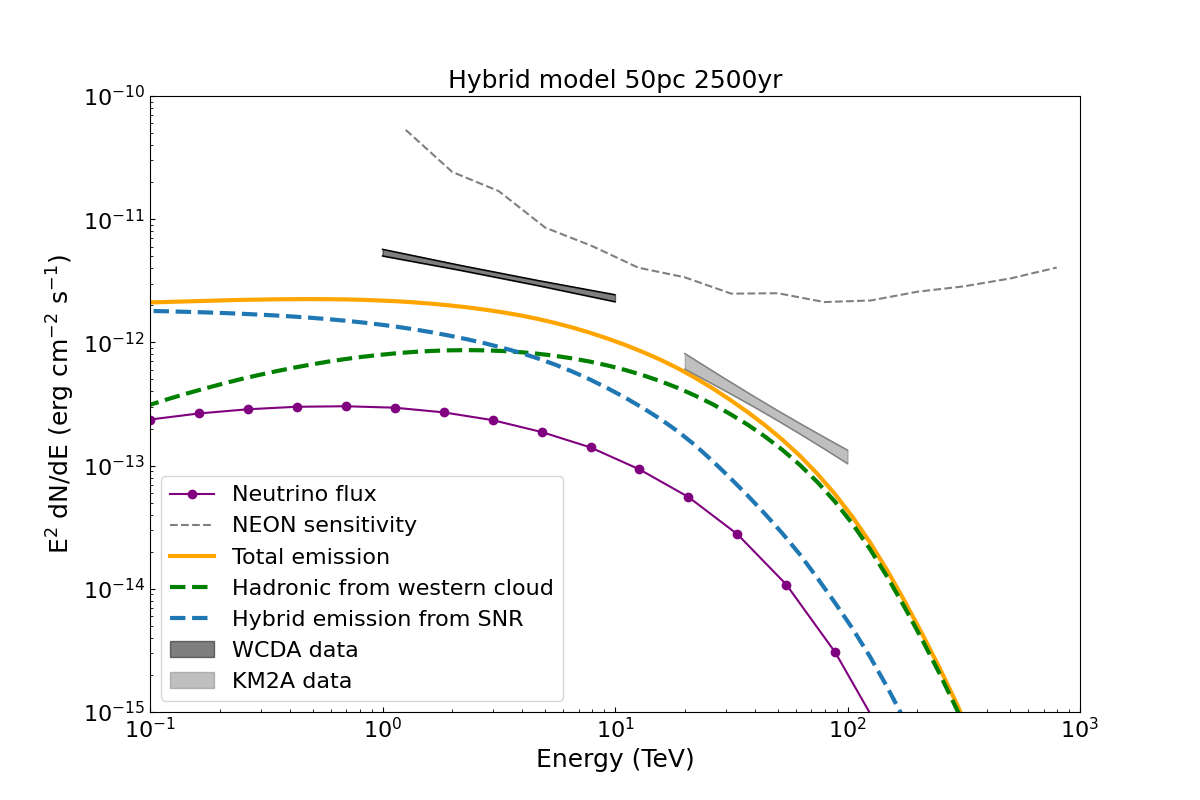}}
    \hfill
    \subfigure[]
    {\includegraphics[width=0.48\textwidth]{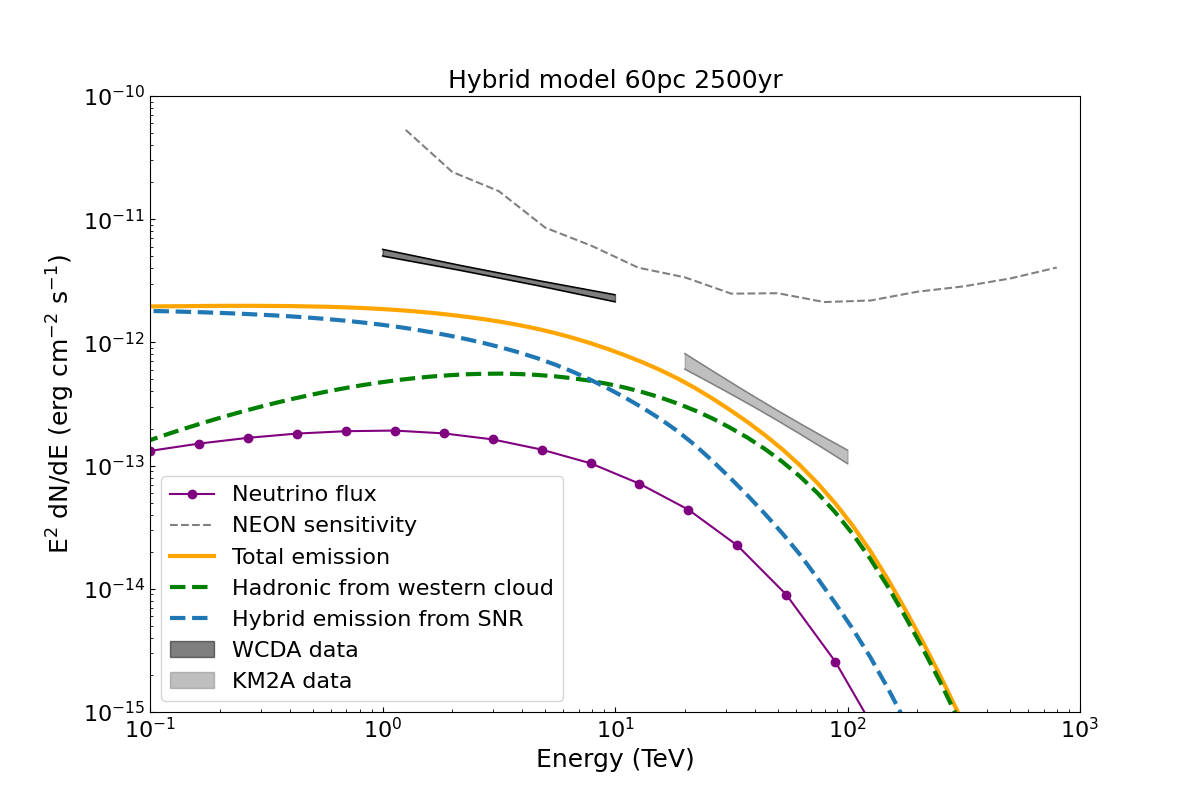}}
    \caption{Hybrid model for different propagation time and distances. The four figures illustrate the best fitting results at various distances. Panel a: diffusion distance 37 pc, average propagation time 1500 years. Panel b: diffusion distance 40 pc, average propagation time 2500 years. Panel c: diffusion distance 50 pc, average propagation time 2500 years. Panel d: diffusion distance 60 pc, average propagation time 2500 years.}
    \label{fig6}
\end{figure}

Moreover, we explore the impact of different propagation distances and average propagation times on the fitting effect as shown in Table \ref{tab:similarity}.
Here the similarity can be defined as $ S = \frac{\sum_{i=1}^{n} S_i}{n}$, where $ s_i =\left| \frac{\text{Fitted data}_i - \text{Observed data}_i}{\text{Observed data}_i} \right| \times 100\% $.
The observed data refer to the WCDA and KM2A data (Figure \ref{fig6}). Specifically, we choose data points of 1, 2.15, 4.64, 10 TeV from WCDA and 39.81, 50.12, 63.10 TeV from KM2A. The fitted data refer to the total emission (Figure \ref{fig6}) and correspond to the same set of energies as those selected for the observed data.

\begin{table}[htbp]
\centering
  \begin{tabular}{|c|c|c|c|c|c|c|c|c|}
  \hline
  \toprule
  \diagbox{D' (pc)}{T (yr)} & 100 & 300 & 500 & 1000 & 1500 & 2000 & 2500 & 2800 \\ \hline
    \midrule

  37(WCDA) & 66\% & 64\% & 67\% & 70\% & 70\% & 70\% & 70\% & 70\% \\ \hline
  37(KM2A) & 89\% & 90\% & 91\% & 88\% & 88\% & 87\% & 87\% & 87\% \\ \hline
  40(WCDA) & 59\% & 57\% & 58\% & 61\% & 61\% & 61\% & 62\% & 62\% \\ \hline
  40(KM2A) & 80\% & 76\% & 82\% & 89\% & 91\% & 91\% & 92\% & 92\% \\ \hline
  50(WCDA) & 45\% & 44\% & 45\% & 46\% & 46\% & 46\% & 46\% & 46\% \\ \hline
  50(KM2A) & 63\% & 59\% & 60\% & 64\% & 66\% & 68\% & 69\% & 68\% \\ \hline
  60(WCDA) & 36\% & 37\% & 37\% & 38\% & 38\% & 39\% & 39\% & 39\% \\ \hline
  60(KM2A) & 60\% & 50\% & 52\% & 54\% & 55\% & 56\% & 57\% & 57\% \\ \hline
  \end{tabular}
\caption{Similarity percentages for different propagation time (T) and distance (D').}
\label{tab:similarity}
\end{table}

When the distance is around 40 pc, the fitting performance for the WCDA data begins to decline, while KM2A still maintains a good fit. However, beyond 50 parsecs, the fitting performance of both WCDA and KM2A noticeably deteriorates. For ultra-long distances, a certain diffusion time is required. When the duration is too long, the fitting effect will decrease. We present the results of the best fit for various distances in Figure \ref{fig6}.

\begin{figure}[htbp]
    \centering
    {\includegraphics[width=0.51\textwidth]{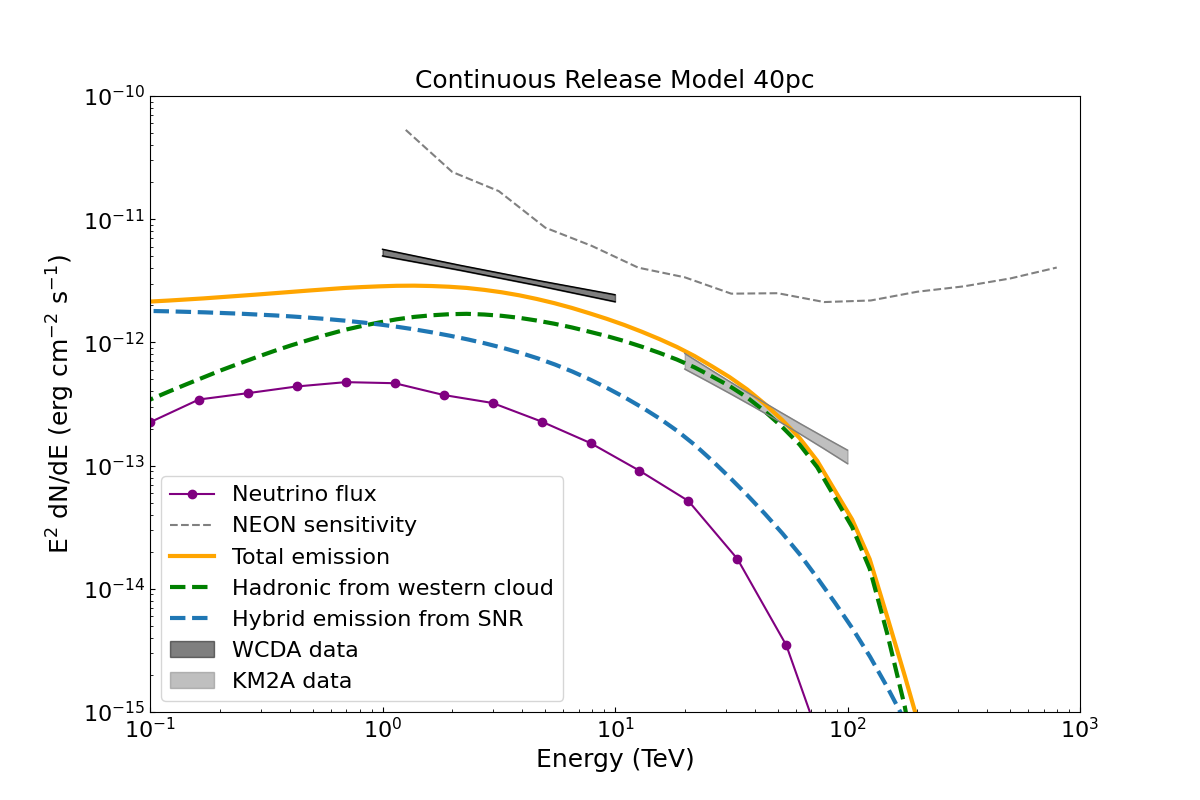}}
    \hfill
    \caption{The result under continuous release model. The emission from the SNR region is labeled as dashed blue line. Hadronic contribution from the Western Cloud is shown as green dashed line. The sum of these two, the total emission, is presented with orange solid line.The predicted neutrino flux are shown as magenta line. }
    \label{fig7}
\end{figure}

Directly using the average release time is a rough approach. We have considered the scenario where particles have been gradually released into interstellar medium. With known SNR size and age we can derive the SNR evolution history according to the Sedov solution. 
Given the shock velocity throughout the entire SNR history, \cite{Kelner:2008ke} provides a rough relationship between $Vshock$ and $Emax$. Here, $Emax$ is the escaped energy of CR. Only when CR with $E > Emax$ they are able to escape from the upstream of SNR shock. Taking the assumption that the total released CR follows a power-law spectrum 
$
 \frac{dN}{dE} = N_p \left( \frac{E}{1 \, \text{Tev}} \right)^{-\alpha} ,
$
we can derive escaped CRs at any time during the SNR history. Here, fitting $\alpha = 1.8$, $Np = 1.63 \times 10^{49}$ corresponds to the total released cosmic-ray energy for 2800 yr. Up to now, the energy released is $1.4 \times 10^{50}$ erg. In this situation, we assume that the SNR is 40 pc away from Earth and at current time the escaped energy is 1 TeV. The best-fit value of the diffusion coefficient is $k = 0.355$ b = 0.39. The fitting results are shown in Figure \ref{fig7}.
Compared to the model with an average propagation time, the continuous propagation model exhibits a slightly better fit.

\vspace{3\baselineskip} 

\subsection{Neutrino flux prediction} \label{3.4}
In the hadronic scenario, neutrinos as counterparts of gamma rays are produced through both $pp$ and $p\gamma$ interactions. Since the cross section of $pp$ interaction is much higher than that of the photohadronic process, here we only consider the scenario of protons colliding with the cold protons in the clouds. Therefore, the neutrino flux can be determined using the following equations \citep{kelner2006energy},

\begin{equation}
\Phi_{\nu_\mu}(E_{\nu_\mu}) = \frac{c n_{MC}}{4\pi d^2} \int \sigma_{pp}\left(\frac{E_{\nu_\mu}}{x}\right) J_p\left(\frac{E_{\nu_\mu}}{x}\right) F_{\nu_\mu}\left(x, \frac{E_{\nu_\mu}}{x}\right) \, \frac{dx}{x},
\label{eq:neutrino_flux}
\end{equation}
where \(x = E_{\nu_\mu} / E_p\) denotes the variable of integration, \(E_{\nu_\mu}\) and \(E_p\) represent the energy of the produced neutrino and the incident proton, respectively. Here \(c\) is the speed of light, \(n_{MC}\) is the density of the molecular cloud, \(d\) is the distance from MC to Earth, and \(J_p\left({E_{\nu_\mu}} / {x}\right)\) denotes the energy distribution of protons as given in Equation \eqref{eq:pp SED}. The inelastic cross section of $pp$ interaction \(\sigma_{pp}\left({E_{\nu_\mu}} / {x}\right)\) can be presented as \citep{kelner2006energy}
\begin{equation}
\sigma_{pp}(E_p) = 34.3 + 1.88L + 0.25L^2 \text{mb},
\label{eq:sigma_pp}
\end{equation}
where $ L = \ln(E_p / 1\text{TeV})$. Muonic neutrinos primarily originate from the decays of pions and muons, hence \(F_{\nu_\mu}\left(x, {E_{\nu_\mu}} / {x}\right)\) consists of two components (\(F_{\nu^{(1)}_\mu},F_{\nu^{(2)}_\mu}\) )\citep{kelner2006energy}, the spectrum of muonic neutrinos produced through the direct decay of pions can be described as follows,

\begin{equation}
F_{\nu^{(1)}_\mu}(x, E_p) = B' \frac{\ln(y)}{y} \left( \frac{1 - y^{\beta'}}{1 + k' y^{\beta'} (1 - y^{\beta'})} \right)^4 \left[ \frac{1}{\ln(y)} - \frac{4 \beta' y^{\beta'}}{1 - y^{\beta'}} - \frac{4 k' \beta' y^{\beta'} (1 - 2 y^{\beta'})}{1 + k' y^{\beta'} (1 - y^{\beta'})} \right],
\label{eq:F_1}
\end{equation}

where $ y = {x}/{0.427} $,

\begin{equation}
B' = 1.75 + 0.204L + 0.010L^2,
\label{eq:B'}
\end{equation}

\begin{equation}
\beta' = \frac{1}{1.67 + 0.111L + 0.0038L^2},
\label{eq:beta'}
\end{equation}

\begin{equation}
k' = 1.07 - 0.086L + 0.002L^2.
\label{eq:k'}
\end{equation}
In the pion decay process, since \(F_{\nu^{(1)}_\mu}\) has a sharp cutoff at $ x = 0.427 $ \citep{kelner2006energy}, the range of integration is set from 0 to 0.427. 
The spectrum of muonic neutrinos produced through the decay of muons can be described as follows,

\begin{equation}
F_{\nu^{(2)}_\mu}(x, E_p) = B'' \frac{\left(1 + k'' (\ln x)^2\right)^3}{x \left(1 + 0.3 / x^{\beta''}\right)} (-\ln(x))^5,
\label{eq:F_2}
\end{equation}

where
\begin{equation}
B'' = \frac{1}{69.5 + 2.65L + 0.3L^2},
\label{eq:B''}
\end{equation}

\begin{equation}
\beta'' = \frac{1}{(0.201 + 0.062L + 0.00042L^2)^{1/4}},
\label{eq:beta''}
\end{equation}

\begin{equation}
k'' = \frac{0.279 + 0.141L + 0.0172L^2}{0.3 + (2.3 + L)^2}.
\label{eq:k''}
\end{equation}
In the muon decay process, the range of integration is set from 0 to 1. The total spectrum of muonic neutrinos is $F_{\nu_\mu}=F_{\nu^{(1)}_\mu}+F_{\nu^{(2)}_\mu}$.
Based on Equations \eqref{eq:neutrino_flux}-\eqref{eq:k''}, we calculated the expected muonic neutrino flux taking into account the neutrino oscillation. 

For the 12-year HESE sample, we take the effective area at 67.6 TeV for this event with a zenith angle of 154.8 degree \citep{IceCube:2013low, Robertson:2019wfw} and calculate the IceCube limit. As seen in Figure \ref{fig5}b, our prediction is consistent with the observation. To finally resolve the mystery of PeVatron, a few next-generation neutrino telescopes have been proposed with enhanced performance, such as IceCube-gen2, P-One, Baikal, TRIDENT, HUNT and NEON \citep{IceCube-Gen2:2021tmd, agostini2020pacific,belolaptikov2021neutrino,TRIDENT:proposal,HUNT:2023mzt}. We show the sensitivity of NEON \citep{Zhang:2024slv} to this source with a spectral index of -2 in Figure \ref{fig5}. The confirmed detection of neutrinos will be able to constraint the hadronic contribution of this source.

\vspace{3\baselineskip} 
\section{Conclusion and discussion} \label{sec:results}
In this article, we attempt to use a hybrid multi-zone scenario to explain the multi-wavelength observation around SNR G54.1+0.3. 
\begin{itemize}
    \item The emission from PWN can very well explain the radio emission and the X-ray emission from PWN, however, it does not explain the GeV to TeV emission due to the high magnetic field inside PWN.
    \item The supernova is very young, using the leptonic emission from the SNR shell can explain well the X-ray emission from the Shell and the GeV to TeV emission from the SNR region. However, LHAASO emission (1TeV-1PeV) is much higher than the VERITAS emission and it covers a much larger era, with the emission center at the Western Cloud.
    \item Assuming that the escaped CRs from the SNR illuminated the Core Cloud and the Western Cloud at $+53 \, \text{km s}^{-1}$, our hybrid model can explain all the observations including those from LHAASO.
\end{itemize}

Our model requires the supernova remnant (SNR) G54.1 + 0.3 capable of accelerating CRs to PeV. 
Theoretically diffusive shock acceleration with magnetic field amplification by non-resonant streaming instability \citep{bell2004turbulent, zirakashvili2008diffusive} can indeed accelerate particles up to ``knee" energies, i.e., up to $\sim10^{15}$ eV.
The key to accelerating particles to PeV energies lies in the presence of strong primordial magnetic field turbulence and high shock velocities. For SNR G54.1+0.3, direct measurement of these parameters is not available. However, with the estimated age of the associated pulsar ($\sim 2900$ years), the observed X-ray shell with a radius of $400''$ and through the Sedov-Taylor solution, we estimate the local ambient density as $ n_H \sim 0.27 \, \text{cm}^{-3} $. This relatively low interstellar medium and the existence of a very young neutron star suggest that SNR G54.1 + 0.3 probably originated from a massive star within a wind-blown bubble. Such an environment often indicates strong magnetic turbulence and high shock velocities in the early stages of SNRs, which makes it possible to accelerate particles up to PeV \citep{zirakashvili2018cosmic}.

Compare to SNR G54.1 + 0.3, G106.3+2.7 \citep{fujita2021x} is a well-known potential PeVatron, whose X-ray and radio observations indicate that it contains a pulsar J2229+6114 and a shell structure. A molecular cloud has also been found in its gamma ray emission region \citep{tibet2021potential}. These conditions are quite similar to those of G54.1+0.3. G106.3+2.7 can be explained with a hybrid model described in Section \ref{sec:model}. However, G106.3 is located only 800 pc from Earth and 6 pc away from the associated molecular cloud. The closer location offers a better fit of the model and a theoretical interpretation.

We expect that future LHAASO observations will be able to use the SNR and the cloud as templates to separate the emission from the clouds and the SNR, providing their spectra.

Our hybrid scenario is based on the association between the SNR and the $+53 \, \text{km s}^{-1}$ clouds, however, no direct evidence of intense collision between the Core Cloud and SNR shock is found. 
Hence, we cannot rule out that LHAASO J1929+1846u is powered from other sources, e.g., another undiscovered PWN/SNR in its vicinity.  We expect future observations on this region by H.E.S.S., CTA, and XMM et al.. Furthermore, we also expect future neutrino telescopes to provide more clues. If the observation of neutrinos from this region is confirmed, the mystery of Galactic PeVatron will be resolved.   

\vspace{3\baselineskip} 

\section*{Acknowledgments}
Many thanks to Caijin Xie, Sujie Lin, Yihan Liu for their suggestions. The authors also thank Songzhan Chen, Xian Hou and Sheng Tang for the helpful discussion.  This work is supported by the National Natural Science Foundation of China (NSFC) grants 12261141691 and the Fundamental Research Funds for the Central Universities, Sun Yat-sen University, No. 24qnpy123. 

\bibliography{reference}{}
\bibliographystyle{aasjournal}



\end{document}